\documentclass[]{article}
\usepackage{amssymb}
\voffset= - 0.3in
\hoffset= - 1.0in         

\catcode`\@=11
\def\marginnote#1{}

\newcount\hour
\newcount\minute
\newtoks\amorpm
\hour=\time\divide\hour by60
\minute=\time{\multiply\hour by60 \global\advance\minute by-\hour}
\edef\standardtime{{\ifnum\hour<12 \global\amorpm={am}%
        \else\global\amorpm={pm}\advance\hour by-12 \fi
        \ifnum\hour=0 \hour=12 \fi
        \number\hour:\ifnum\minute<10 0\fi\number\minute\the\amorpm}}
\edef\militarytime{\number\hour:\ifnum\minute<10 0\fi\number\minute}

\def\draftlabel#1{{\@bsphack\if@filesw {\let\thepage\relax
   \xdef\@gtempa{\write\@auxout{\string
      \newlabel{#1}{{\@currentlabel}{\thepage}}}}}\@gtempa
   \if@nobreak \ifvmode\nobreak\fi\fi\fi\@esphack}
        \gdef\@eqnlabel{#1}}
\def\@eqnlabel{}
\def\@vacuum{}
\def\draftmarginnote#1{\marginpar{\raggedright\scriptsize\tt#1}}

\def\draft{\oddsidemargin -.5truein
        \def\@oddfoot{\sl preliminary draft \hfil
        \rm\thepage\hfil\sl\today\quad\militarytime}
        \let\@evenfoot\@oddfoot \overfullrule 3pt
        \let\label=\draftlabel
        \let\marginnote=\draftmarginnote
   \def\@eqnnum{(\theequation)\rlap{\kern\marginparsep\tt\@eqnlabel}%
\global\let\@eqnlabel\@vacuum}  }

\def\d{\partial}

\def\bea{\begin{eqnarray}}
\def\eea{\end{eqnarray}}

\def\beq{\begin{equation}}
\def\eeq{\end{equation}}
\def\ba{\beq\new\begin{array}{c}}
\def\ea{\end{array}\eeq}
\def\be{\ba}
\def\ee{\ea}
\def\stackreb#1#2{\mathrel{\mathop{#2}\limits_{#1}}}
\def\Tr{{\rm Tr}}
\def\F{{\cal F}}
\makeatletter
\newdimen\normalarrayskip              
\newdimen\minarrayskip                 
\normalarrayskip\baselineskip
\minarrayskip\jot
\newif\ifold             \oldtrue            \def\new{\oldfalse}
\def\arraymode{\ifold\relax\else\fi} 
\def\eqnumphantom{\phantom{(\theequation)}}     
\def\@arrayskip{\ifold\baselineskip\z@\lineskip\z@
     \else
     \baselineskip\minarrayskip\lineskip2\minarrayskip\fi}
\def\@arrayclassz{\ifcase \@lastchclass \@acolampacol \or
\@ampacol \or \or \or \@addamp \or
   \@acolampacol \or \@firstampfalse \@acol \fi
\edef\@preamble{\@preamble
  \ifcase \@chnum
     \hfil$\relax\arraymode\@sharp$\hfil
     \or $\relax\arraymode\@sharp$\hfil
     \or \hfil$\relax\arraymode\@sharp$\fi}}
\def\@array[#1]#2{\setbox\@arstrutbox=\hbox{\vrule
     height\arraystretch \ht\strutbox
     depth\arraystretch \dp\strutbox
     width\z@}\@mkpream{#2}\edef\@preamble{\halign
\noexpand\@halignto
\bgroup \tabskip\z@ \@arstrut \@preamble \tabskip\z@ \cr}%
\let\@startpbox\@@startpbox \let\@endpbox\@@endpbox
  \if #1t\vtop \else \if#1b\vbox \else \vcenter \fi\fi
  \bgroup \let\par\relax
  \let\@sharp##\let\protect\relax
  \@arrayskip\@preamble}
\def\eqnarray{\stepcounter{equation}%
              \let\@currentlabel=\theequation
              \global\@eqnswtrue
              \global\@eqcnt\z@
              \tabskip\@centering
              \let\\=\@eqncr
 \halign to \displaywidth\bgroup
    \eqnumphantom\@eqnsel\hskip\@centering
    $ \tabskip\z@ {##}$%
    \global\@eqcnt\@ne \hskip 2\arraycolsep
         $\arraymode{##}$\hfil
    \global\@eqcnt\tw@ \hskip 2\arraycolsep
         $\tabskip\z@{##}$\hfil
         \tabskip\@centering
    &{##}\tabskip\z@\cr}
\begingroup\ifx\undefined\newsymbol \else\def\input#1 {\endgroup}\fi
\newfont{\hr}{msbm10}
\newfont{\ams}{msam10}

\font\numbers=cmss12
\font\upright=cmu10 scaled\magstep1
\def\stroke{\vrule height8pt width0.4pt depth-0.1pt}
\def\topfleck{\vrule height8pt width0.5pt depth-5.9pt}
\def\botfleck{\vrule height2pt width0.5pt depth0.1pt}
\def\Zmath{\vcenter{\hbox{\numbers\rlap{\rlap{Z}\kern 0.8pt\topfleck}\kern
2.2pt
                   \rlap Z\kern 6pt\botfleck\kern 1pt}}}
\def\Qmath{\vcenter{\hbox{\upright\rlap{\rlap{Q}\kern
                   3.8pt\stroke}\phantom{Q}}}}
\def\Nmath{\vcenter{\hbox{\upright\rlap{I}\kern 1.7pt N}}}
\def\Cmath{\vcenter{\hbox{\upright\rlap{\rlap{C}\kern
                   3.8pt\stroke}\phantom{C}}}}
\def\Rmath{\vcenter{\hbox{\upright\rlap{I}\kern 1.7pt R}}}
\def\Z{\ifmmode\Zmath\else$\Zmath$\fi}
\def\Q{\ifmmode\Qmath\else$\Qmath$\fi}
\def\N{\ifmmode\Nmath\else$\Nmath$\fi}
\def\C{\ifmmode\Cmath\else$\Cmath$\fi}
\def\R{\ifmmode\Rmath\else$\Rmath$\fi}

\def\stackreb#1#2{\mathrel{\mathop{#2}\limits_{#1}}}
\def\Tr{{\rm Tr}}
\def\res{{\rm res}}

\def\d{\partial}

\def\half{{\textstyle{1\over2}}}
\def\2{{1\over 2}}
\def\N2{${\cal N}=2$}
\def\4N{${\cal N}=4$}
\def\1N{${\cal N}=1$}
\def\1N*{${\cal N}=1^*$}

\def\beq{\begin{equation}}
\def\eeq{\end{equation}}
\def\ba{\beq\new\begin{array}{c}}
\def\ea{\end{array}\eeq}
\def\be{\ba}
\def\ee{\ea}
\def\stackreb#1#2{\mathrel{\mathop{#2}\limits_{#1}}}
\newcommand{\rf}[1]{(\ref{#1})}

\begin{document}


\begin{flushright}
FIAN/TD-06/05\\
ITEP/TH-22/05
\end{flushright}


\begin{center}
\baselineskip20pt
{\bf \LARGE Matrix Model and Stationary Problem in Toda Chain
\footnote{Based on talks at "Classical and quantum integrable systems", Dubna,
January 2005 and "Selected topics of modern mathematical physics", St.Petersburg,
June 2005, and a lecture for the minicourse: "Toda lattices: basics and perspectives",
Fields Institute, Toronto, April 2005}}
\end{center}
\bigskip
\begin{center}
\baselineskip12pt
{\large A.~Marshakov}\\
\medskip
{\em Lebedev Physics Institute and ITEP\\
Moscow, Russia}\\
{\sf e-mail:\ mars@lpi.ru, mars@itep.ru}\\
\end{center}
\bigskip\bigskip\medskip

\begin{center}
{\large\bf Abstract} \vspace*{.2cm}
\end{center}

\begin{quotation}
\noindent
We analyze the stationary problem for the Toda chain, and show that arising
geometric data exactly correspond to the multi-support solutions of
one-matrix model with a polynomial potential. For the first nontrivial
examples the Hamiltonians and symplectic forms are calculated explicitly, and
the consistency checks are performed. The corresponding quantum problem is formulated
and some its properties and perspectives are discussed.
\end{quotation}



\setcounter{equation}{0}

\section{Introduction}

It was found long ago (see, for example \cite{Migdal,David})
and emphasized recently \cite{CIV,DV} in the context of string geometry
and its certain consequences for the
supersymmetric gauge theories, that the solution to one-matrix model
\be
\label{mamo}
Z = \int_{N\times N} d\Phi \exp\left(-{1\over\hbar}\Tr W(\Phi)\right)
\ee
with the polynomial potential
\be
\label{mamopot}
W(\Phi) = \sum_{k=1}^{n+1} t_k\Phi^k
\ee
in planar approximation within the $1/N$-expansion
is geometrically described in terms of the hyperelliptic complex curve
\be
\label{mamocu}
y^2 = W'(\lambda)^2 - 4f_{n-1}(\lambda)
\ee
endowed with generating differential (related to the eigenvalue density for the
matrix $\Phi$)
\be
\label{mamodiff}
dS = yd\lambda
\ee
The parameters of potential \rf{mamopot} in this context are certain fixed constants
or "Casimirs", while the dynamical variables are rather related with nontrivial periods
of generating differential \rf{mamodiff} on hyperelliptic curve \rf{mamocu}.

This geometric setup implies, in particular, that the most effective language for
the planar solution of \rf{mamo} is theory of integrable systems. Basically, already in
\cite{David} it was found that the partition function \rf{mamo} $\F = \hbar^2\log Z$ at
$N\to\infty$ with fixed $\hbar N$ is geometrically formulated as a (logarithm of)
quasiclassical tau-function \cite{KriW}. This is a particular case of the so called
algebro-geometric "holomorphic" integrable systems, quite commonly arising recently in
the context of string theory or Seiberg-Witten theory (see e.g. \cite{Mbook,HKbook} and
references therein).

On the other hand,
it is also well-known that the matrix integral \rf{mamo}, as a function of parameters
of the potential \rf{mamopot} and the size of the matrix $N$, is a tau-function of the
(semi-infinite, forced) Toda chain \cite{GMMMO}. The Toda chains are well-known families
of integrable models
arising in particular context of the algebro-geometric integrable systems: their affine
or periodic versions are literally the Seiberg-Witten integrable systems associated to
the pure (\N2 SUSY) gauge theories \cite{GKMMM}. One may be interested then, whether
the integrable system, associated to \rf{mamocu}, \rf{mamodiff} has any relation to the
Toda family.

Below we are going to demonstrate, that the complex curve
\rf{mamocu} with the differential \rf{mamodiff} is directly associated
with a particular integrable
Toda system, namely with a {\em stationary} problem in the Toda
chain
\footnote{Analogous stationary problem for the KdV hierarchy is well-known and
widely discussed in the literature, see \cite{SKdV}.}. We discuss the general structure
of the classical stationary problem, present explicit
computations for the simplest nontrivial examples where the "order $n$ is $n=2$ and $n=3$, and discuss briefly
the problems of its quantization. We hope that the observed equivalence of the matrix model
and stationary Toda geometries will be first step towards understanding of proper
analogs of the extra quasiclassical variables, and the considered below system can
play the role of the corresponding "probe" model.

\section{Notations}

The Toda chain Lax operator is an infinite 3-diagonal matrix
\be
\label{todalax}
L_{ij} = r_i\delta_{j,i-1} + p_i\delta_{ji} + r_{i+1}\delta_{j,i+1}
\ee
(with $i,j\in\mathbb{Z}$), where it is convenient to introduce
\be
\label{r}
r_i=e^{\2(q_i-q_{i-1})}
\ee
or
\be
\label{R}
R_i=r_i^2=e^{q_i-q_{i-1}}
\ee
to be useful in what follows.

The Toda chain equations of motion follow from the Lax representation
\be
\label{laxeq}
{\d L\over\d t_1} = \left[ L,A_1 \right]
\ee
with $A_1=\half {\cal R}\circ L$ or, explicitly for the ${\cal R}$-matrix
\be
\label{a1}
\left( A_1\right)_{ij} = \2\left(r_{i+1}\delta_{j,i+1}-r_i\delta_{j,i-1}\right)
\ee
providing \rf{laxeq} being equivalent to the well-known equations of motion
\be
\label{todaeq}
{\d r_i\over\d t_1} = r_i\left(p_i-p_{i-1}\right)
\\
{\d p_i\over\d t_1} = r_{i+1}^2-r_i^2 = R_{i+1}-R_i
\ee
for the chain of particles with the nearest neighbor exponential interaction.

In a standard way, the nonlinear equations \rf{laxeq} can be formulated as consistency
of auxiliary linear problem
\be
\label{auxili}
\left( L\psi\right)_i = r_{i+1}\psi_{i+1} + p_i\psi_i + r_i\psi_{i-1} = \lambda\psi_i
\ee
The basis of two-dimensional space of solutions to the linear problem
can be found in the form of expansion
\be
\label{psipm}
\psi^{\pm}_i = \lambda^{\pm i}e^{\mp\2 q_i}\left(1+{\xi^\pm_i\over\lambda}+
{\eta^\pm_i\over\lambda^2}+\dots\right)
\ee
where for the coefficients of expansion one gets
\be
\label{kpm}
\xi^+_{i-1}-\xi^+_i = p_{i-1}
\\
\xi^-_i-\xi^-_{i-1} = p_i
\ee
and
\be
\label{epm}
\eta^+_{i-1}-\eta^+_i = R_{i-1}+ \xi^+_{i-1}(\xi^+_{i-1}-\xi^+_i)
\\
\eta^-_i-\eta^-_{i-1} = R_{i+1}+ \xi^-_i(\xi^-_i-\xi^-_{i-1})
\ee
which relates them directly with the Toda chain phase space variables.

\section{Stationary problem}

An infinite Toda chain is a completely integrable system with (infinitely
many) conserving charges, which generate the "higher flows", commuting with \rf{laxeq}.
The higher flows can be written as
\be
\label{laxhigh}
{\d L\over\d t_k} = \left[ L,A_k \right]
\ee
with the same ${\cal R}$-matrix $A_k = \half {\cal R}\circ L^k$, as in \rf{a1}.
There are many ways to restrict the infinite Toda chain to a
finite-dimensional subspace of its phase space; most well-known are $N$-periodic
chain and open chain (the "Toda molecule") with $N$ particles.

Below we are going to concentrate on the stationary problem, determined by
commutativity condition
\be
\label{stat}
\left[ L,A \right]=0
\ee
with
\be
\label{a}
A = \sum_{k=1}^n c_k A_k
\ee
for some fixed number - "order" $n$, and some coefficients $\{ c_k\}$.

As any problem with commutativity of two operators, the
stationary problem \rf{stat} acquires a complex curve
\be
\label{cudet}
\det\left(A(\lambda)-Y\right)=0
\ee
where matrix
\be
\label{amatrix}
A(\lambda) = \left(
\begin{array}{cc}
  a(\lambda) & b(\lambda) \\
  c(\lambda) & d(\lambda) \\
\end{array}\right)
\ee
with some polynomial matrix elements
is representation of the operator \rf{a} in the basis of two-dimensional space of the
eigenfunctions \rf {auxili} of the operator \rf{todalax}.
The curve \rf{cudet} is therefore hyperelliptic
\be
\label{curve}
y^2 = P_{2n}(\lambda) \equiv \left(a-d\right)^2+4bc
\ee
with $y=2Y-\Tr A(\lambda)=2Y-a-d$.

The most natural symplectic form for the infinite Toda chain is just
$\Omega_\infty=\sum_{i\in\mathbb{Z}} \delta q_i\wedge\delta p_i$. It is also known
that it "remains intact", when reduced to the $2N$-dimensional phase space of the
$N$-periodic problem or $N$-particle open molecule, i.e.
$\Omega_N = \sum_{i=1}^N \delta q_i\wedge\delta p_i$. The latter one can be also
written as $\Omega_N = \res_{w=\infty} \left({dw\over w}\ \Tr\left(\Psi^{-1}\delta L(w)
\wedge\delta\Psi\right)\right)$ in terms of the spectral parameter dependent Lax
operator of the $N$-periodic problem, see \cite{KriPho,Msf,Mbook} for details.

Similarly,
the symplectic structure for the stationary problem can be determined by
\be
\label{sform}
\Omega = \res_{\lambda=\infty} \left(d\lambda\ \Tr\left(\Psi^{-1}\delta A(\lambda)
\wedge\delta\Psi\right)\right)
\ee
where $\delta A(\lambda)$ is now the variation (at constant spectral parameter
$\lambda$!) of the operator \rf{a}, \rf{amatrix} and
\be
\label{psim}
\Psi = \left(
\begin{array}{cc}
  \psi^+_i & \psi^-_i \\
  \psi^+_{i-1} & \psi^-_{i-1} \\
\end{array}\right)
\ee
is the matrix of the Baker-Akhiezer functions, with $\Psi^{-1}$ being its
matrix inverse.

In what follows we will use for convenience
a different from \rf{psipm} normalization of the Baker-Akhiezer functions:
\be
\label{psimatr}
\Psi \rightarrow \Psi \left(
\begin{array}{cc}
  e^{q_i\over 2} & 0 \\
  0 & e^{-{q_{i-1}\over 2}} \\
\end{array}\right)
\ee
Note, that the normalization of the Baker-Akhiezer function
does not influence the symplectic form \rf{sform}, since
the difference, say for the change of normalization \rf{psimatr}, is proportional to
\be
\res \left(\delta Y d\lambda\right)\wedge \delta(q_i-q_{i-1}) = 0
\ee

\section{Explicit computations}

Let us now illustrate the general construction by explicit computations.

\subsection{n=1}

This is the trivial case of stationarity of the operator \rf{a1}, which in the basis of
eigenfunctions \rf{auxili} reads
\be
\label{a1l}
A_1(\lambda)\left(
\begin{array}{c}
  \psi_i \\
  \psi_{i-1} \\
\end{array}\right) = \left(
\begin{array}{cc}
  \half(\lambda-p_i) & -r_i \\
  r_i & -\half(\lambda-p_{i-1}) \\
\end{array}\right)\left(
\begin{array}{c}
  \psi_i \\
  \psi_{i-1} \\
\end{array}\right)
\ee
giving rise to the genus zero curve \rf{curve}
\be
\label{curven1}
y^2=\lambda^2-(p_i+p_{i-1})\lambda+{1\over 4}(p_i+p_{i-1})^2+4R_i
\ee
with the expansion
\be
\label{ydx1}
yd\lambda\ \stackreb{\lambda\to\infty}{\sim}\ d\lambda\left(\lambda -
\2(p_i+p_{i-1})+{2R_i\over\lambda} + \dots
\right)
\ee
The commutativity condition \rf{stat} $[L,A_1]=0$ leads to the further constraints
\be
\label{cc1}
p_{i+1}=p_i=p_{i-1}
\\
R_{i+1}=R_i
\ee
and fixing the coefficients of \rf{curven1} or \rf{ydx1} means that all variables
turn into some constants.

In the trivial $n=1$ case, explicit calculation of \rf{sform} gives
\be
\label{sfn1}
\Omega_1 = e^{q_i-q_{i-1}}dq_i\wedge dq_{i-1}+
\2\left(d\xi_i^+\wedge d\xi_i^- + d\xi_{i-1}^+\wedge d\xi_{i-1}^-\right)
-d\xi_i^+\wedge d\xi_{i-1}^-
\ee
Since
\be
\2\left(d\xi_i^+\wedge d\xi_i^- + d\xi_{i-1}^+\wedge d\xi_{i-1}^-\right)
-d\xi_i^+\wedge d\xi_{i-1}^-\ \stackreb{\rf{kpm}}{=}\ \2\left(d\xi_i^+\wedge dp_i
+ dp_{i-1}\wedge d\xi_{i-1}^-\right)
\ee
and due to \rf{cc1} the symplectic form \rf{sfn1} vanishes.

\subsection{n=2}

Consider now the simplest nontrivial case of $c_k \propto \delta_{kn}$ for $n=2$, i.e.
$A=A_2 = \half {\cal R}\circ L^2$. Applying twice \rf{todalax}, one gets
\be
\label{l2}
\left( L^2\psi\right)_i = r_{i+2}r_{i+1}\psi_{i+2} +
r_{i+1}\left(p_{i+1}+p_i\right)\psi_{i+1} + \left(R_{i+1}+p_i^2+R_i\right)\psi_i
+ r_i\left(p_i+p_{i-1}\right)\psi_{i-1} + r_ir_{i-1}\psi_{i-2}
\ee
or
\be
\label{a2}
\left( A_2\psi\right)_i \propto r_{i+2}r_{i+1}\psi_{i+2} +
r_{i+1}\left(p_{i+1}+p_i\right)\psi_{i+1} -
r_i\left(p_i+p_{i-1}\right)\psi_{i-1} - r_ir_{i-1}\psi_{i-2}
\ee
while stationarity of the gradient of the trace part
$\Tr L^2 = \sum_i\left(p_i^2+2R_i\right)$ in \rf{l2},
which disappears from \rf{a2}, lead
exactly to the constraints \rf{cc1} of the previous trivial example.

The commutativity constraints lead now to
\be
\label{cc2}
R_{i+1}(p_{i+1}+p_i) = R_i(p_i+p_{i-1})
\\
R_{i+1}+p_i^2=R_{i-1}+p_{i-1}^2
\ee
which reduce the dimension of the phase space to the four independent
variables ($R_i$ and $p_i$ on two neighboring sites).

Now, using \rf{auxili}, one can rewrite \rf{a2} in the basis of eigenfunctions of
the Lax operator \rf{todalax}, e.g.
\be
\label{a2l}
A_2(\lambda)\left(
\begin{array}{c}
  \psi_i \\
  \psi_{i-1} \\
\end{array}\right) = \left(
\begin{array}{cc}
  \half(\lambda^2-p_i^2-R_{i+1}+R_i) & -r_i(\lambda+p_i) \\
  r_i(\lambda+p_{i-1}) & -\half(\lambda^2-p_{i-1}^2-R_{i-1}+R_i) \\
\end{array}\right)\left(
\begin{array}{c}
  \psi_i \\
  \psi_{i-1} \\
\end{array}\right)
\ee
In particular, upon \rf{cc2} one gets
$\Tr A_2(\lambda)\ \stackreb{\rf{cc2}}{=}\ 0$.

For the operator \rf{a2l} the equation of the curve \rf{curve} reads
\be
\label{curven2}
y^2 = \left(\lambda^2-\half Q\right)^2-4R_i(\lambda+p_i)(\lambda+p_{i-1})
\\
Q \equiv R_{i+1}+R_{i-1}-2R_i+p_i^2+p_{i-1}^2
\ee
The expansion of the generating differential at $\lambda\to\infty$ gives
\be
\label{ydx2expan}
yd\lambda\ \stackreb{\lambda\to\infty}{\sim}\ d\lambda\left(\lambda^2
-\2(Q+4R_i) - {2R_i\over\lambda}(p_i+p_{i-1})-{R_i\over\lambda^2}
(2p_ip_{i-1}+Q+2R_i) +  \dots\right) = \\
\ \stackreb{Q+4R_i=2C_2}{=}\ d\lambda\left(\lambda^2
-C_2 - {2C_1\over\lambda}-{2R_i\over\lambda^2}
\left(p_ip_{i-1}-R_i+C_2\right) - {2C_1C_2\over\lambda^3}
-{2\over\lambda^4}(C_1^2+C_2H)
+  \dots\right)
\ee
where fixing $2C_2=R_{i+1}+R_{i-1}+2R_i+p_i^2+p_{i-1}^2$ together with the coefficient
at ${d\lambda\over\lambda}$ lead to the following Casimir constraints
\be
\label{c2}
C_1=R_i(p_i+p_{i-1})
\\
C_2=R_{i+1}+R_i+p_i^2=R_{i-1}+R_i+p_{i-1}^2
\ee
which, in addition to \rf{cc2} reduce the phase space to be two-dimensional. A
convenient choice of co-ordinates is $R_i$ (for some fixed site $i$) and the
difference between the corresponding momenta $p_-=p_i-p_{i-1}$.

The Hamiltonian $H$ is, up to a numeric factor, a coefficient at the
${d\lambda\over\lambda^2}$-term in \rf{ydx2expan}
\be
\label{hamn2}
H = R_i\left(p_ip_{i-1}-R_i+C_2\right)\ \stackreb{p_i-p_{i-1}\equiv p_-}{=}\
{C_1^2\over 4}{1\over R_i} - {1\over 4}R_ip_-^2 -R_i + C_2R_i
\ee
With formulas \rf{c2} equation of the curve \rf{curven2} acquires the form
\be
\label{cun2}
y^2 = (\lambda^2 - C_2)^2 - 4(C_1\lambda + H)
\ee
and coincides exactly with the matrix model curve \rf{mamocu} for the potential
$W = {\lambda^3\over 3}-C_2\lambda$ and function $f_1=C_1\lambda + H$.

Explicit calculation of the symplectic form for the $n=2$ case, using \rf{psimatr}
and \rf{amatrix}, gives
\be
\label{o2}
\Omega_2
=\half d\left(R_{i-1}+R_i+p_{i-1}^2\right)\wedge d\xi_{i-1}^- -
\half d\left(R_{i+1}+R_i+p_i^2\right)\wedge d\xi_i^+ +
d(p_i-p_{i-1})\wedge dR_i =
\\
\stackreb{\rf{c2}}{=}
d(p_i-p_{i-1})\wedge dR_i
\ee
Now, in order to fit between the hamiltonian \rf{hamn2}, the curve \rf{cun2}
and the symplectic form \rf{o2} one should rewrite equation \rf{cun2}
in the "Seiberg-Witten" form (see e.g. \cite{2tc} and \cite{Mbook}), i.e. as
\be
\label{cun2w}
w + {C_1\lambda + H\over w} = \lambda^2 - C_2
\\
w - {C_1\lambda + H\over w} = y
\ee
This means, in particular, that $2w = y+\lambda^2-C_2$, and
\be
\label{sfyw}
\half dy\wedge d\lambda = dw\wedge d\lambda
\ee
The first of the equations \rf{cun2w} can now be written as
\be
\label{Hw}
H = w\lambda^2 - C_1\lambda - C_2w - w^2 = w\left(\lambda-{C_1\over 2w}\right)^2
- {C_1^2\over 4w} - C_2w - w^2
\ee
which exactly coincides with \rf{hamn2}, provided by the identifications
\be
\label{idn2}
w \leftrightarrow -R_i
\\
\lambda-{C_1\over 2w} \leftrightarrow \pm p_- 
\ee

\subsection{n=3}

For the $n=3$, with  $c_k \propto \delta_{k3}$ the explicit expression for the operator
$A_3 = \half {\cal R}\circ L^3$ looks like
\be
\label{a3}
\left( A_3\psi\right)_i \propto r_{i+3}r_{i+2}r_{i+1}\psi_{i+3}+
r_{i+2}r_{i+1}(p_{i+2}+p_{i+1}+p_i)\psi_{i+2} + \\ +
r_{i+1}\left(p_{i+1}^2+p_{i+1}p_i+p_i^2+r_{i+2}^2+R_{i+1}+R_i\right)\psi_{i+1} - \\ -
r_i\left(p_i^2+p_ip_{i-1}+p_{i-1}^2+R_{i+1}+R_i+R_{i-1}\right)\psi_{i-1} - \\-
r_{i-1}r_{i-2}(p_i+p_{i-1}+p_{i-2})\psi_{i-2}-r_ir_{i-1}r_{i-2}\psi_{i-3}
\ee
while stationarity of the gradient of the trace part
\be
\label{trl3}
{1\over 3}\Tr L^3 = {1\over 3}\sum_{i} p_i^3 + \sum_i p_i(R_i+R_{i+1})
\ee
gives rise to the constraints \rf{cc2} of the previous example.
The commutativity conditions \rf{stat} now give
\be
\label{coc3}
p_i^3+\left(p_{i+1}+2p_i\right)R_{i+1}+p_iR_i =
p_{i-1}^3+\left(p_{i-2}+2p_{i-1}\right)R_{i-1}+p_{i-1}R_i
\\
R_{i+1}\left(p_{i+1}^2+p_{i+1}p_i+p_i^2+R_{i+2}+R_{i+1}+R_i\right) =
R_i\left(p_i^2+p_ip_{i-1}+p_{i-1}^2+R_{i+1}+R_i+R_{i-1}\right)
\ee
In the basis of eigenfunctions of the Lax operator
matrix $A_3(\lambda)$ for \rf{a3} acquires the form
\be
\label{a3l}
\left(
\begin{array}{cc}
  \half(\lambda^3+2r_i^2\lambda-p_i^3-(p_{i+1}+2p_i)r_{i+1}^2+p_{i-1}r_i^2) &
-r_i(\lambda^2+p_i\lambda+p_i^2+r_{i+1}^2+r_i^2) \\
  r_i(\lambda^2+p_{i-1}\lambda + p_{i-1}^2+r_i^2+r_{i-1}^2) &
-\half(\lambda^3+2r_i^2\lambda-p_{i-1}^3-(p_{i-2}+2p_{i-1})r_{i-1}^2+p_ir_i^2) \\
\end{array}\right)
\ee
For the operator \rf{a3l} the curve \rf{curve} reads
\be
\label{curven3}
y^2 =
\lambda^{6} -\left(p_i^{3} + p_{i-1}^{3} + (p_{i-2}+2p_{i-1})R_{i-1} +
3(p_i+ p_{i-1})R_i
 + (p_{i+1} + 2 p_i)R_{i+1} \right)\lambda^{3} -  \\
-4R_i\left(R_{i+1} + R_i + R_{i-1} + p_i^{2} + p_ip_{i-1} +
p_{i-1}^{2}\right)\lambda^{2} - 2R_i\left(
(p_{i+1} + 2p_i + 2p_{i-1})R_{i+1}+\right. \\ \left.  +  (p_i+p_{i-1})R_i +
(p_{i-2}+2p_{i-1} + 2p_i)R_{i-1} +p_i^{3}  + 2p_i^{2}p_{i-1}
 + 2p_ip_{i-1}^{2} + p_{i-1}^{3}
\right)\lambda \\
 +  {1\over 4}\left((p_{i+1}+2p_i)R_{i+1}
- (p_i+p_{i-1})R_i +(p_{i-2}+2p_{i-1})R_{i-1}+p_i^{3}+ p_{i-1}^{3}\right)^{2} - \\
 - 4R_i\left( p_i^{2} + R_{i+1} +R_i\right)\left(p_{i-1}^{2}+ R_i +
R_{i-1}\right)
\ee
The expansion of the generating differential is
\be
\label{ydx3expan}
yd\lambda\ \stackreb{\lambda\to\infty}{\sim}\ d\lambda\left(\lambda^3
-\half\left((p_{i+1}+2p_i)R_{i+1} + 3(p_i+p_{i-1})R_i +
(p_{i-2}+2p_{i-1})R_{i-1} + p_i^{3} + p_{i-1}^{3}\right) - \right. \\
\left. -
{2R_i\over\lambda}\left(R_{i+1} + R_i + R_{i-1} + p_i^{2} + p_ip_{i-1} +
p_{i-1}^{2}\right)-
{R_i\over\lambda^2}\left(
(p_{i+1}+ 2p_i+ 2p_{i-1})R_{i+1} +\right.\right. \\  \left.\left.+
(p_i + p_{i-1})R_i +
(p_{i-2}+2p_{i-1}+ 2p_i)R_{i-1} + p_i^{3}  + 2p_i^{2}p_{i-1}
+ 2p_ip_{i-1}^{2} + p_{i-1}^{3} \right) -\right. \\ \left. -
{R_i\over\lambda^3}\left( (p_{i+1}p_i + p_{i+1}p_{i-1} +2p_i^{2} +2p_ip_{i-1} +
 2p_{i-1}^{2} )R_{i+1} + (3p_i^{2} +3p_{i-1}^{2} + 2p_ip_{i-1} )R_i+
\right.\right. \\  \left.\left.+
(p_{i-1}p_{i-2} + p_ip_{i-2} + 2p_ip_{i-1} +
2p_{i-1}^{2}  + 2p_i^{2})R_{i-1}  +
2(R_{i+1}R_i+R_{i+1}R_{i-1}+R_i^2+R_iR_{i-1})+
\right.\right. \\  \left.\left.+
p_i^{4} + p_i^{3}p_{i-1} + 2p_i^{2}p_{i-1}^{2} + p_ip_{i-1}^{3} + p_{i-1}^{4}
\right) + \dots \right)
\ee
Fixing the coefficients at singular terms of the expansion of $yd\lambda$
to be the independent upon dynamical variables
numbers (or "Casimirs"), using \rf{a3l}, one can write
\be
\label{cc3}
C_1 = R_i\left(p_i^2+p_ip_{i-1}+p_{i-1}^2+R_{i+1}+R_i+R_{i-1}\right)
\\
C_3=p_i^3+\left(p_{i+1}+2p_i\right)R_{i+1}+\left(p_{i-1}+2p_i\right)R_{i}=
p_{i-1}^3+\left(p_{i-2}+2p_{i-1}\right)R_{i-1}+\left(2p_{i-1}+p_i\right)R_{i}
\ee
Together with \rf{coc3}, relations \rf{cc3} impose four constraints to the
eight-dimensional space of variables ($R_j$ and $p_j$ with $j=i-1,i,i+1$ together with
$R_{i-2}$ and $p_{i+2}$), so that the phase space in this example is four-dimensional.

Then the expansion
\rf{ydx3expan} upon \rf{cc3} turns into
\be
\label{ydx3e}
yd\lambda\ \stackreb{\lambda\to\infty}{\sim}\ d\lambda\left(\lambda^3- C_3 -
{2R_i\over\lambda}\left(R_{i+1} + R_i + R_{i-1} + p_i^2 + p_ip_{i-1} +
p_{i-1}^2 \right)- \right. \\ \left. -
{2R_i\over\lambda^2}\left(p_{i-1}R_{i+1} - (p_i + p_{i-1})R_i +
p_iR_{i-1} + p_i^2p_{i-1} + p_ip_{i-1}^2+C_3\right) + \dots
\right) =
\\
= d\lambda\left(\lambda^3-C_3 - {2C_1\over\lambda} - {2H_1\over\lambda^2} - {2H_2\over\lambda^3} + \dots\right)
\ee
so that the first of two independent hamiltonians (proportional to the
coefficients at ${d\lambda\over\lambda^2}$ and ${d\lambda\over\lambda^3}$) can be, using \rf{cc3}, presented as
\be
\label{h31}
H_1 = C_3R_i + {C_1\over 2}(p_i+p_{i-1})-\2 R_i\left(p_i^3+p_{i-1}^3+
3R_i(p_i+p_{i-1})+(p_i-p_{i-1})(R_{i+1}-R_{i-1})\right)
\ee
and
\be
\label{h32}
H_2={C_1^2\over 4R_i} + {C_1\over 2}\left(R_i-p_ip_{i-1}\right) + C_3R_i(p_i+p_{i-1})
+{1\over 4}R_i^3 -
\\
-{1\over 4}R_i(R_{i+1}-R_{i-1})^2 - \2 R_i(R_{i+1}-R_{i-1})(p_i^2
-p_{i-1}^2) -
\\
- {1\over 4}R_i\left(p_i^4+3p_i^2p_{i-1}^2+p_{i-1}^4\right)
- R_i^2\left(
p_i^2+{5\over 3}p_ip_{i-1}+p_{i-1}^2\right)
\ee
The equation of the curve \rf{curven3} now becomes
\be
\label{cu3}
y^2 = (\lambda^3-C_3)^2 - 4(C_1\lambda^2 +H_1\lambda + H_2)
\ee
again coinciding with \rf{mamocu} with the potential $W = {\lambda^4\over 4}-C_3\lambda$
and function $f_2 = C_1\lambda^2 +H_1\lambda + H_2$.

Explicit calculation of the symplectic form $\Omega_3$ gives
\be
\label{o3draft}
\Omega_3 =
 \2 d\xi_{i-1}^- \wedge d\left(
p_{i-1}^3 +(p_{i-2}+2p_{i-1})R_{i-1}+(p_i+2p_{i-1})R_i
\right) + \\
+ \2 d\left(
p_i^3 +(p_{i+1}+2p_i)R_{i+1}+(p_{i-1}+2p_i)R_i
\right)\wedge d\xi_i^+
\\
+ d(R_{i+1}-R_{i-1})\wedge dR_i + 2R_idp_i\wedge dp_{i-1} +
\\
+ d\left(p_i^2-p_{i-1}^2\right)\wedge dR_i +
\left(p_{i-1}dp_i-p_idp_{i-1}\right)\wedge dR_i
\ee
Using \rf{cc3}, and introducing $R_-\equiv R_{i+1}-R_{i-1}$ this slightly simplifies to
\be
\label{o3}
\Omega_3 =
dR_-\wedge dR_i + 2R_idp_i\wedge dp_{i-1} +
\\
+ d\left(p_i^2-p_{i-1}^2\right)\wedge dR_i +
\left(p_{i-1}dp_i-p_idp_{i-1}\right)\wedge dR_i
\ee
(certainly, $d(\Omega_3)=0$), or
\be
\Omega_3 = \left(dR_-\ dR_i\ dp_i\ dp_{i-1}\right)\ \cdot
\hat\Omega_3\ \cdot
\left(\begin{array}{c}
  dR_- \\
  dR_i \\
  dp_i \\
  dp_{i-1} \\
\end{array}\right)
\ee
where the wedge product is implied and $\hat\Omega_3$ is the matrix
\be
\label{o3ma}
\hat\Omega_3 = \left(
\begin{array}{cccc}
0 & \frac{1}{2}  & 0 & 0 \\
-\frac{1}{2} & 0 &  - \left(p_i+\frac{p_{i-1}}{2}\right)  & p_{i-1} +
\frac {p_i}{2}  \\
0 & p_i + \frac {p_{i-1}}{2}  & 0 & R_i \\
0 &  - \left(p_{i-1}+\frac{p_i}{2}\right)  &
 - R_i & 0
\end{array} \right)
\ee
A nontrivial check is that the Hamiltonians \rf{h31} and \rf{h32}
indeed commute
\be
\label{hcom3}
\{ H_1, H_2 \}_{\Omega_3} = 0
\ee
with respect to the Poisson bracket, corresponding to symplectic structure \rf{o3} or
defined by the inverse to \rf{o3ma} matrix
\be
\label{o3po}
\left(
\begin{array}{crcc}
0 & -2 & \frac{2p_{i-1} + p_i}{R_i}
 & \frac{2p_i + p_{i-1}}{R_i}  \\
2 & 0 & 0 & 0 \\
 - \frac{2p_{i-1} + p_i}{R_i}  & 0
 & 0 &  -  \frac{1}{R_i}  \\  - \frac{2p_i + p_{i-1}}{R_i}  & 0
 & \frac{1}{R_i}  & 0
\end{array} \right)
\ee
giving rise to the "elementary brackets"
\be
\label{elpo3}
\{ R_i,R_- \}_{\Omega_3} = 1
\\
\{ p_i,p_{i-1} \}_{\Omega_3} = - {1\over 2R_i}
\\
\{ R_-,p_i\}_{\Omega_3} = {1\over R_i}\left(p_{i-1}+{p_i\over 2}\right)
\\
\{ R_-,p_{i-1}\}_{\Omega_3} = {1\over R_i}\left(p_i+{p_{i-1}\over 2}\right)
\ee

The symplectic structure \rf{o3} can be brought to canonical form,
introducing new variables as $p_i=p\ \cosh\theta$ and
$p_{i-1}=p\ \sinh\theta$, then
\be
\label{o3theta}
\Omega_3 = dR_-\wedge dR_i + R_id(p^2)\wedge d\theta +
p^2 dR_i\wedge d\theta + d(p^2)\wedge dR_i =
\\
= d(R_- + p^2)\wedge dR_i + d(R_ip^2)\wedge d\theta
\ee
and one concludes that the Darboux co-ordinates are $\theta$, $R\equiv R_i$
together with
\be
\label{rd}
\rho = R_ip^2
\\
\Delta = R_- + p^2
\ee
In these variables the Poisson brackets are
\be
\{ \Delta, R \}_{\Omega_3} = 1, \ \ \ \ \{ \rho, \theta \}_{\Omega_3} = 1
\ee
while all other vanish. The Hamiltonians \rf{h31}, \rf{h32}
in these variables have the form
\be
\label{H3D}
H_1
= C_3R + \frac {C_1}{2}\rho^{1/2}R^{-1/2} e^{\theta}
- \frac {1}{2}\Delta\rho^{1/2}R^{1/2}e^{-\theta}
+ \frac {1}{8}\left( e^{-\theta}-
e^{3\theta}\right)\rho^{3/2}R^{-1/2}
   - \frac {3}{2}\rho^{1/2}R^{3/2} e^{\theta}
\\
H_2 = \frac {C_1^{2}}{4R} + \frac {C_1}{2}R
 + \frac{C_1}{8}{\rho\over R}\left(e^{-2\theta}-e^{2\theta}\right) +
C_3\rho^{1/2}R^{1/2}e^{\theta} + \frac {1}{4}R^{3}
- \frac {1}{4} \Delta^{2}R - \frac {1}{32} {\rho^{2}\over R}
- \frac {9}{8}\rho R e^{2\theta} +
\\
+ \frac {1}{8} \rho Re^{-2\theta}   +
 \frac {1}{64} \frac {\rho^{2}}{R}\left(e^{4\theta} + e^{-4\theta}\right)
\ee
i.e. are generally functions of the
{\em fractional} powers of dynamical variables. The dependence on
fractional powers, however, disappears for $H_2$ at vanishing Casimir $C_3=0$ and
for the square $H_1^2$ at $C_1=C_3=0$.

\section{Discussion}

Let us, finally, discuss some related issues and open problems.

\paragraph{Geometry on stationary and periodic problems}\

When defining the stationary problem, we have started with an infinite Toda chain.
Remaining all ingredients almost intact, one could take as initial point a $N$-periodic
problem in Toda instead ($q_{i+N}=q_i$, $p_{i+N}=p_i$, $\psi_{i+N} = w\psi_i$ etc) with
sufficiently large $N$.

Such procedure, however, may be already treated purely in terms of algebro-geometric
integrable systems. One may start with the curve of $N$-periodic chain
\be
\label{NToda}
w + {1\over w} = P_N(\lambda)
\ee
endowed with generating differential $dS = \lambda{dw\over w}$ and consider its
reduction down to $2n$-dimensional subspace in the moduli space for $n<N$. A natural way
to impose such reduction is to require existence on the curve \rf{NToda}
of a single-valued meromorphic function with
only two poles (at $\lambda=\infty$) of order $n$. Existence of such function
on the curve \rf{NToda} is
exactly equivalent to stationarity of the order $n$ flow in periodic Toda, associated
with the linear combination of meromorphic differentials
$d\Omega_A=\sum_{k=1}^n c_kd\Omega_k$.
Since there is no natural function with desired properties on the curve
\rf{NToda} in general position, this constraint effectively reduces the (smooth) genus
of the Riemann surface
from $N-1$ to $n-1$ and one ends up with the curve of a stationary Toda chain
\rf{mamocu}. Such reduction was already discussed in the literature comparing the
Seiberg-Witten and Dijkgraaf-Vafa geometries, see \cite{CIV} and, for example,
\cite{Hol,Bo,ItoKa}. Note also, that the corresponding generating differentials
or symplectic structures remain unrelated by this procedure, and this exactly
corresponds to the difference between the symplectic structures \rf{sform},
\rf{o2} and \rf{o3} of stationary problem and that of the
infinite chain we discussed above.

\paragraph{Remarks on quantum case}\

For the curve \rf{cun2} and the symplectic form \rf{sfyw} one can
immediately write the "naive" Schr\"odinger equation
\be
\label{cuschro}
\hbar^2{\d^2\over\d\lambda^2} \Upsilon = \left(W'(\lambda)^2-
\hbar W''(\lambda)\right)\Upsilon
\ee
with the solution
\be
\label{soschro}
\Upsilon = \exp\left(-{1\over\hbar}W(\lambda)\right)
\ee
since
\be
\label{decomp}
\hbar^2{\d^2\over\d\lambda^2} - W'(\lambda)^2 + \hbar W''(\lambda)
= \left(\hbar{\d\over\d\lambda}-W'(\lambda)\right)
\left(\hbar{\d\over\d\lambda}+W'(\lambda)\right)
\ee
Comparison of \rf{cuschro} with \rf{cun2} for the potential
$W'(\lambda) = \lambda^2-C_2$ gives $H=0$ and $C_1={\hbar\over 2}$, and
comparison with \rf{cu3} leads to $H_2=H_1=0$ and $C_1={3\over 4}\hbar$.
Formulas \rf{cun2w} and \rf{sfyw} rather suggest that the quantization
can be better performed in the $(y,w)$-variables. For the periodic Toda chain
\rf{NToda} this way leads to the Baxter second-order difference equation, but for
the stationary problem equation \rf{cun2w}, due to \rf{sfyw} would give rise to an
{\em integral} operator. The main obstacle on this way is that it is vague what
are the "proper" variables for quantization of stationary problem, since using the naive
Darbough co-ordinates, already in $n=3$ case, leads to the non-analytic
dependencies in the Hamiltonians.

One can nevertheless try to impose the quantum relations
\be
\label{HZ}
\hat {\cal H}_i {\cal Z} = H_i {\cal Z},\ \ \ i=1,\ldots,n-1
\ee
onto the matrix model ${\cal Z}$-function, which instead of naive partition function
$Z$ defined by integral representation \rf{mamo} contains explicitly information about
(quantum version of) geometry \rf{mamocu}, \rf{mamodiff}, i.e. one literally has to
solve \rf{HZ} for the function
\be
\label{cZ}
{\cal Z}(c,H) = \exp\left({1\over\hbar^2}F_0(c,H)+\ldots\right)
\ee
This formulation of quantum problem is consistent with
the fact that the form of Hamiltonians \rf{H3D} simplifies at vanishing (clasically)
Casimirs, when they are of order of $\hbar $ or being some quantum corrections. The
classical formulas \rf{H3D} for $H_i$ (or some polynomial functions of them)
are then replaced by differential and difference
operators $\hat {\cal H}_i$ via $\Delta\rightarrow {\d/\d R}$ and
$\theta\rightarrow{\d/\d\rho}$, so that $e^{\pm\theta}$ turn into the shift operators
in $\rho$-variable. Constructed in this way ${\cal Z}$-function \rf{cZ}
can be treated as a matrix-model analogue of Nekrasov's generalization of the
Seiberg-Witten prepotential \cite{NekrasovSW}.

\section*{Acknowledgements}

I am grateful to H.~Braden, R.~Donagi, H.~Kanno, S.~Kharchev, B.~Khesin,
A.~Mironov, N.~Nekrasov, M.~Olshanetsky, V.~Rubtsov,
A.~Veselov and, especially, to I.~Krichever for very useful discussions.
The work was partially supported by the RFBR grant 04-01-00642, the
grant for support of scientific schools 1578.2003.2, the Federal
Program of the Russian Ministry of Industry, Science and Technology
No 40.052.1.1.1112, and the Russian Science Support Foundation.

\end{document}